\documentclass[preprint,showpacs,preprintnumbers,amsmath,amssymb]{revtex4}
\begin{document}
\title{New Results in Axion Physics}

\author{\underline{\textrm{Eduard Mass{\'o}}} and Javier Redondo}

\affiliation{Grup de F{\'\i}sica Te{\`o}rica and Institut de
F{\'\i}sica d'Altes Energies,
\\Universitat Aut{\`o}noma de Barcelona,\\
08193 Bellaterra, Barcelona, Spain}


\baselineskip=11.6pt

\begin{abstract}
  We comment on the recent experimental results from the PVLAS and
CAST collaborations that search for axion-like particles. We propose
a particle physics model in which their apparent inconsistency is
circumvented.
\end{abstract}

\maketitle

\section{Introduction}
The search for new very weakly interacting light particles was
motivated by the theoretical idea of the axion, a particle
arising\cite{Weinberg:1977ma,Wilczek:1977pj} when one introduces the
Peccei-Quinn symmetry\cite{Peccei:1977ur,Peccei:1977hh} in order to
solve the strong CP-problem. When allowing the breaking scale of the
symmetry to be a very high energy scale one obtains the so called
invisible axion, named this way because of its very weakly
interactions. Also, the theory of the axion predicts a very small
mass for it. A milestone in axion physics was the realization by
Sikivie that it was not impossible to probe the existence of the
axion with feasible experiments\cite{Sikivie:1983ip}.

In\cite{Sikivie:1983ip} there are some proposals to look for axions.
All are based on the axion-photon conversion in a magnetic field, an
effect that is described by the interaction term
\begin{equation}
{\cal
L}_{\phi\gamma\gamma}=\frac{1}{4M}F^{\mu\nu}\widetilde{F}_{\mu\nu}\phi
\ \ \ \ \ . \label{L}
\end{equation}
Here $F^{\mu\nu}$ is the electromagnetic field tensor,
$\widetilde{F}_{\mu\nu}$ its dual, and $\phi$ the axion field. One
may envisage to detect axions if they constitute the galactic dark
matter halo (haloscope). Another type of experiment (helioscope) is
based on the fact that light particles should be produced in the
interior of the Sun, and subsequently leave it in the form of a
continuous flux. At the Earth, we can try to detect this solar flux
by looking at axion conversion to X-rays in a magnetic field.
Finally, there is the remarkable effect of light shining through a
wall: in a magnetic field, light oscillates into axions, these cross
a wall and afterwards they convert back into photons.

Here we will be concerned with helioscopes and also with another
detection method that was proposed in\cite{Maiani:1986md}. It
consists in letting a polarized laser to propagate in a magnetic
field. The real conversion of photons to axions produces a selective
absorption of one of the two laser polarization (dichroism), and the
virtual conversion photon-axion-photon produces a phase retardation
of one of the polarizations (birefringence). If we write  (\ref{L})
as
\begin{equation}
{\cal
L}_{\phi\gamma\gamma}=\frac{1}{M}\overrightarrow{E}\cdot\overrightarrow{B}
\phi
\end{equation}
we see that it is the polarization (taken as the direction of the
electric field in the laser) parallel to the external magnetic field
that gets affected.

Let us mention that any light particle coupled to two photons could
be able to give a positive signal in these experiments. This general
situation has been studied in\cite{Masso:1995tw}. In the case that
the particle is a pseudoscalar, we would have a Lagrangian similar
to (\ref{L}). Notice that a light scalar particle, with  a
lagrangian given by
\begin{equation}
{\cal L}'_{\phi\gamma\gamma}=\frac{1}{4M}F^{\mu\nu}F_{\mu\nu}\phi
\label{LS}
\end{equation}
also may be detected in an helioscope, may induce light shining
through walls and also may be responsible of dichroism and
birefringence, although it is now the polarization perpendicular to
the magnetic field that intervenes, since (\ref{LS}) can be written
as
\begin{equation}
{\cal L}'_{\phi\gamma\gamma}=\frac{1}{2M} \left (
\overrightarrow{E}^2 - \overrightarrow{B}^2 \right) \phi \ .
\end{equation} I will refer to these hypothetical particles, either
pseudoscalar or scalar, as axion-like particles.

Recently, there have been two groups that have announced
experimental results on axion-like particles coupled to photons.
First, there is the helioscope of the CAST collaboration that has
not observed any signal coming from the Sun and has put the
limit\cite{Zioutas:2004hi}
\begin{equation}
M > 0.9 \times 10^{10}  \ {\rm GeV}   \label{cast}
\end{equation}
valid for  a mass of the light particle $m < 0.02$ eV.

The second result is by the PVLAS collaboration. They have a
positive signal
 consistent in a rotation of the plane of the polarization of a laser
 propagating in a magnetic field\cite{Zavattini:2005tm}. Their result can be interpreted
 in terms of an axion-like particle. It is consistent with a scale
\begin{equation}
M \sim 4 \times 10^{5}\ {\rm GeV}  \label{pvlas}
\end{equation}
and with a light particle mass $m \sim 10^{-3}$ eV.

We notice that the two results are in strong contradiction. In fact
the PVLAS result is also in contradiction with the stellar energy
loss bounds. These bounds are obtained when noticing that light
$\phi$ particles would be produced by the Primakoff-like effect due
to the interaction (\ref{L}) or (\ref{LS}). If the produced $\phi$'s
escape freely from the star the whole process constitutes a
non-standard channel of energy-loss, which is limited by
observational data on stellar evolution time scales. These arguments
lead to the bound\cite{Raffelt:1999tx}
\begin{equation}
M >  10^{10} \ {\rm GeV} \label{stellar} \ .
\end{equation}

If all these results are confirmed of course we will need to find a
model where we can understand (\ref{pvlas}), (\ref{cast}), and
(\ref{stellar}) at the same time.

\section{Our Model}

Let us start discussing two possible ways to evade (\ref{stellar})
 as discussed in\cite{Masso:2005ym}. First, we may assume that $\phi$'s
are indeed produced in the Sun core, but they get trapped because of
some additional interactions and diffuse with extremely small mean
free path up to the solar surface. This could in principle evade the
astrophysical bounds. Unfortunately a satisfactory model has yet to
be found\cite{Masso:2005ym,Jain:2005nh}. A second possibility is
that there is much less production in the Sun than expected. We
shall here report recent work we have done\cite{Masso2006} along
this second line. For related approaches see\cite{Jaeckel:2006id}.

\begin{figure}[h]
  \vspace{3cm}
  \includegraphics{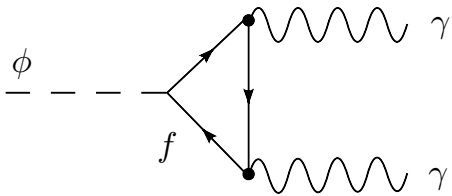}
  \caption{\it
    Triangle diagram for the $\phi\gamma\gamma$ vertex.
   \label{fig1} }
\end{figure}

We shall assume that the neutral $\phi$ particle couples to two
photons through a triangle with a new fermion $f$; see
Fig.(\ref{fig1}).

We wish that this new particle $f$ has small electric charge on the
one hand, and on the other hand we want this charge to decrease when
going from the momentum transfer involved in the PVLAS experiment,
$|k^2|\, \sim $ 0, to the typical momentum transfer in the solar
processes, $|k^2|\, \sim$ keV$^2$. In order for this second
condition to happen it is already clear that we need to advocate for
new physics with a very low energy scale.

We can meet both conditions in the context of paraphoton
models\cite{Okun:1982xi,Holdom}. To start with, these models are the
only ones, as far as we know, where the effective electric charge of
some particles can be naturally very small. The idea is that
particles with paracharge get an induced electric charge
proportional to some small mixing angle $\epsilon$ between photons
and paraphotons. To satisfy the second condition, i.e., to get a
variation of the effective electric charge with energy, we use a
model with two light but massive paraphotons with the same mixing
with the photon. If the fermion $f$ couples to the two paraphotons
with opposite paracharge, the resulting effective electric charge
for $f$ decreases with energy or temperature $T$,
\begin{equation}
q_f(T) \simeq \frac{\mu^2}{T^2} q_f(0) \label{decrease}
\end{equation}
where $\mu$ is the mass scale of the paraphoton masses (in
(\ref{decrease}) we assume $T>>\mu$). With $\mu \simeq 10^{-3}$ eV
and $\epsilon$ such that $q(0)e \simeq 10^{-8}e$, our model is able
to accommodate the strength of the PVLAS signal and yet have a very
suppressed emission in the Sun. Notice that in our model the CAST
limit (\ref{cast}), which is based on a standard solar $\phi$-flux,
does not hold.

Summarizing, we have presented a paraphoton model that evades the
astrophysical bounds on axion-like particles and is consistent with
the CAST and the PVLAS experimental results.

\section{Acknowledgements}

We would like to congratulate Mario Greco as well as Giorgio
Bellettini and Giorgio Chiarelli for the superb organization of the
XXth session of the workshop. EM thanks Alvaro De R{\'u}jula for
asking him the right question in the workshop. We acknowledge
support by the CICYT Research Project FPA2005-05904 and the
\textit{Departament d'Universitats, Recerca i Societat de la
Informaci{\'o}} (DURSI), Project 2005SGR00916.

\end{document}